# Yields of Projectile Fragments in Sulphur-Emulsion Interactions at 3.7A GeV

S. Kamel [1;1)], W. Osman [2;2)] and M. Fayed [2;3)]

[1] Physics Department, Faculty of Education, Ain Shams University, Cairo11757, Egypt

[2] Physics Department, Faculty of Science, Cairo University, Giza 12613, Egypt

**Abstract:** This work presents the basic characteristics of singly, doubly and heavily charged projectile fragments (PFs) emitted in inelastic interactions of $^{32}$S ions with photo-emulsion nuclei at Dubna energy (3.7A GeV). Our experimental data are compared with the corresponding data for other projectiles at the same incident energy. The study of mean multiplicities of different charged PFs against the projectile mass shows a power-law relationship. The multiplicity distributions of singly and doubly charged PFs have been fitted well with a Gaussian distribution function. The yields of PFs broken up from the interactions of $^{32}$S projectile nuclei with different target nuclei are studied. The beam energy dependence in terms of the various order moments is studied as well.
.

## 1 Introduction

Over the past three decades the study of relativistic nucleus–nucleus interactions has been very important, especially at energies higher than 1A GeV. As a result of interactions between the individual nucleons of colliding nuclei, a lot of energy is available. This energy is transformed into plentiful production of pions. On a much longer time scale, de–excitation of the residual projectile and target nuclei occurs, which is called the nuclear multifragmentation process [1]. An incident projectile nucleus collides with a target nucleus, then the nonviolent collision parts of the projectile and target nuclei multifragment into some smaller fragments [2].

---

Corresponding author   1) S. Kamel sayedks@windowslive.com

Authors             2) wafaabahr61@gmail.com  3) maha_fayed@hotmail.com

The phenomenon of multifragmentation is important particularly for heavy nuclei because a phase transition from liquid to gas is expected to happen in the non-violent collision part. However, in the violent collision part, a quark–gluon plasma is expected to appear. Considerable achievements have been made in the study of nuclear multifragmentation in heavy–ion collisions at Dubna energy, and have attracted much interest [3-13].

We focus in the present work on projectile fragmentation in the interactions of 3.7A GeV sulphur ions with emulsion nuclei and target groups (light and heavy) of nuclei. Additionally, we compare our findings with the corresponding results from other labs for different projectiles at the same Dubna energy. Finally, we investigate the dependence of the fragmentation processes on sulphur ion energy at the lowest (Dubna, 3.7A GeV) and highest (CERN, 200A GeV) available energies in terms of the various order moments.

## 2  Experimental Details

In the present experiment, a nuclear emulsion stack of the type NIKFI-BR2 was exposed to $^{32}$S beams of energy 3.7A GeV at the JINR Synchrophasotron in Dubna, Russia. Every pellicle of the emulsion has dimensions of 20 cm x 10 cm x 600 µm. The pellicles were scanned under 100 × magnification with an "along-the-track" technique. Depending on the particle ionization, all the charged secondary particles (tracks) have been classified according to the commonly accepted emulsion terminology [14, 15] as:

1- Shower particles: They mainly consist of pions. Their multiplicity is denoted by $n_s$.
2- Gray and black particles (target fragments): Their multiplicities are denoted $n_g$ and $n_b$, respectively. The heavily ionizing particle multiplicity denoted $n_h$ (= $n_g + n_b$) gives good estimation of the excitation energy of the target nucleus.
3- The non-interacting projectile fragments (projectile spectators) are highly collimated tracks within a narrow forward cone of opening angle θ ≤ 3° for $^{32}$S at 3.7A GeV. These projectile fragments (PFs) are classified according to their charges into:

   i)   Singly charged PFs (Z = 1): their multiplicity is denoted by $N_P$. From Ref. [10], the relativistic hydrogen isotopes protons, deuterons and tritons are produced in nuclear emulsion by the ratio 77.6%, 19.1% and 3.3%, respectively.

ii) Doubly charged PFs (Z = 2): their multiplicity is denoted by $N_{alpha}$. It is reported in Ref. [10] that the helium isotopes $^4$He and $^3$He are emitted in nuclear emulsion by the ratios 76.3% and 23.7%, respectively.

iii) Multiply charged PFs (Z ≥ 3): their multiplicity is denoted by $N_F$. They indicate no change in their ionization up to a distance of at least 1cm from the interaction vertex. Their charge can be determined by the δ-ray counting method [16].

The experimental qualitative measure is usually chosen to be either the heavy-ionizing-particle multiplicity $n_h$ (which is a target size parameter) or the total charge of the projectile spectators *Q* or both. The quantity *Q* is a suitable experimental parameter to classify nucleus–nucleus interactions and the degree of their peripherality, such that interactions with small *Q* are considered central collisions with a low impact parameter *b*, and events with large *Q* are peripheral collisions with a large *b*.

It is known that nuclear emulsion is a heterogeneous target composed of hydrogen (H, $A_T$ =1), light (CNO, $A_T$ =14), and heavy (AgBr, $A_T$ = 94) nuclei. The separation of events into ensembles of collisions of sulphur projectiles with H, CNO and AgBr nuclei needs an event-by-event analysis method. There are several methods to differentiate events according to these target groups, as detailed in experiments [17, 18]. Usually events with $n_h \leq 1$ are classified as collisions with hydrogen. Events with $2 \leq n_h \leq 7$ are classified as CNO events. Events with $n_h \geq 8$ arise from collisions with heavy nuclei. In this method, the separation of events for an AgBr target is quite accurate in samples with $n_h \geq 8$, but in samples with $2 \leq n_h \leq 7$, there is an admixture of peripheral collisions with the AgBr target. In samples with $n_h \leq 1$, there is an admixture of peripheral collisions with the CNO and AgBr targets. In fact, we divide the events into three samples: $n_h \leq 1$, $2 \leq n_h \leq 7$ and $n_h \geq 8$, which are represented by H, CNO and AgBr, respectively.

## 3  Results and discussion

### 3.1  Dependence of the PFs on projectile size

In nuclear reactions at relativistic and ultra-relativistic energies, when the products of interest are projectile-like and target-like fragments, the dominant reaction mechanism at the relativistic energies is represented by the geometric spectator-participant model, where a hot region is formed in the participant zone (zone of geometric overlap), while the spectator regions are colder and can survive the process of de-excitation as projectile-like fragments. Still, some of these spectators can be warm enough to undergo multifragmentation, especially in semi-central collisions.

In this work the multiplicity distributions (MDs) of singly Z = 1, doubly Z = 2 and multiply Z ≥ 3 charged PFs for $^{32}$S-Em interactions at 3.7A GeV are shown in Fig. 1(a, b and c). The corresponding data for different projectiles [7] ($^{4}$He, $^{12}$C, $^{16}$O, $^{22}$Ne and $^{28}$Si) at ~ 3.7A GeV are illustrated in the same figure. From this figure, the width of the MD of Z= 1 (Fig. (1a)) is extended to more values than that for the MD of Z = 2 (Fig. 1(b)) and for the MD of Z ≥ 3 (Fig. 1(c)) as well, suggesting the validity of the limiting fragmentation hypothesis (LFH). In Fig. 1(b), for all the projectiles, the height of the distribution of doubly charged PFs ($N_{alpha}$) decreases with increasing number of emitted alpha PFs. Besides, the production of doubly charged PFs is dependent on the mass of the incident projectile. In the present interaction, the production probability of one He isotope is around 47% of the total events; in which helium products are present, and it is the dominant process in all the collisions considered. The production probabilities of two and three He isotopes per event are about 35% and 18%, respectively [12]. In Fig. 1(c), the MD of multiply charged PFs ($N_F$) shows a decrease in the peak value with increasing projectile mass, except for the $^{28}$Si projectile. The charge spectrum of Z ≥ 3 PF emitted in $^{32}$S-Em interactions is inserted on the right side of Fig. 1(c). Our findings show very few events that have more than one PF with charge Z ≥ 3. This comment is compatible with the results of the probabilities for the different fragmentation modes studied in Ref. [16] by one of our authors. Table 1 displays these probabilities for sulphur fragmentation from nuclear peripheral interactions in emulsion at 3.7 and 200A GeV.

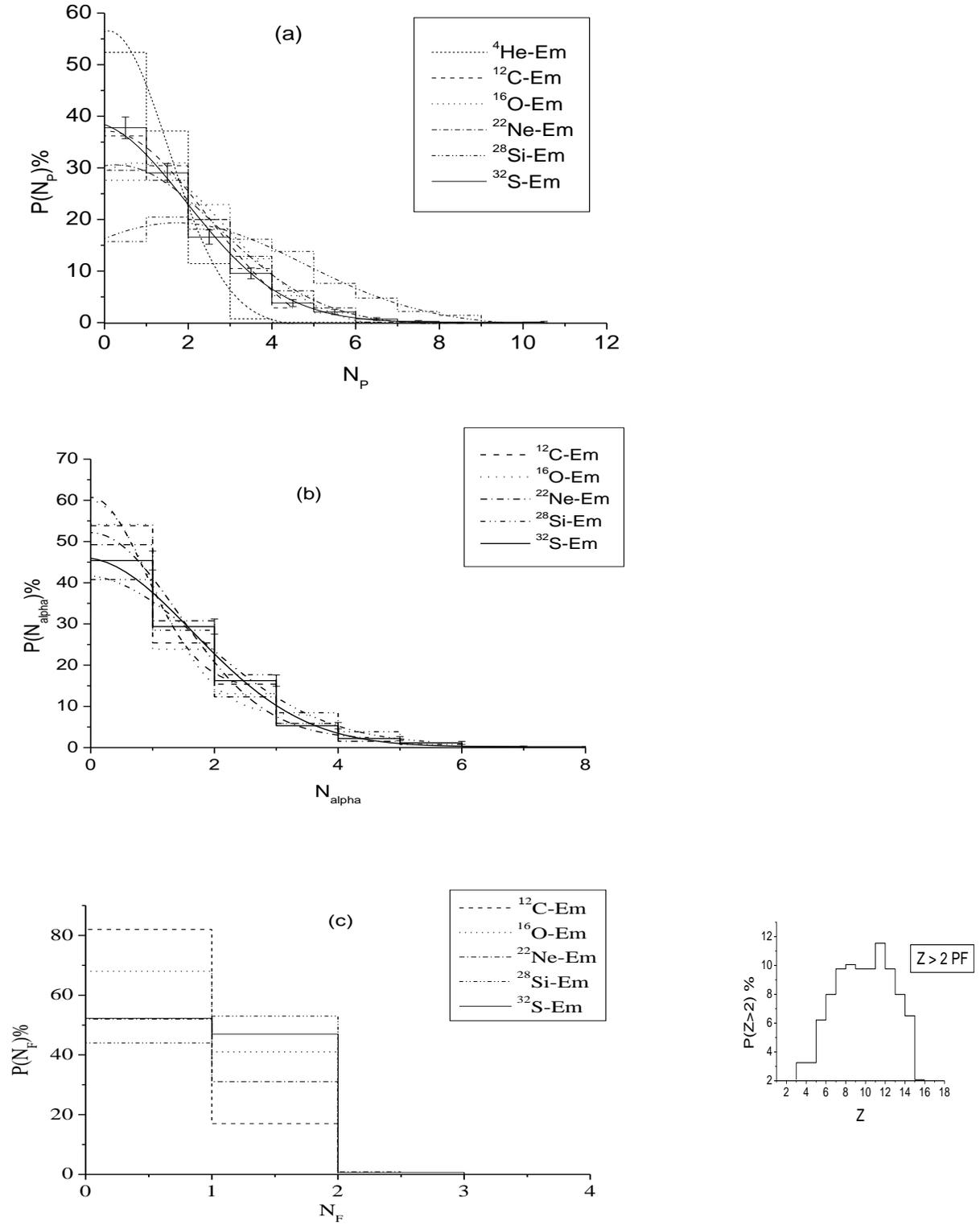

**Fig. 1.** Multiplicity distributions of a) singly, b) doubly and c) multiply charged PFs in nucleus-emulsion interactions at ~ 3.7A GeV. The charge spectrum of Z > 2

PF emitted in $^{32}$S-Em interactions is inserted on the right-hand side of (c). The data in (a) and (b) are fitted with a Gaussian distribution.

**Table 1.** Some parameters describing the sulphur fragmentation in emulsion at 3.7 and 200A GeV (see Ref. 16).

| Fragmentation mode | Probability (%) for different fragmentation modes | |
|---|---|---|
| | 3.7A GeV | 200A GeV |
| Two fragments with $Z > 2$ | 2.50 ± 1.40 | 2.24 ± 0.53 |
| One fragment with $Z > 2$ | 87.50 ± 8.50 | 62.76 ± 2.80 |
| Fragments with $Z > 2$ and no $\alpha$ particles | 30.80 ± 5.10 | 31.34 ± 2.80 |
| Fragments with $Z > 2$ and $\alpha$ particle | 59.20 ± 7.00 | 33.66 ± 2.90 |
| $\alpha$ particle and no heavier fragments | 10.00 ± 2.89 | 34.69 ± 2.94 |

The mean values of singly $<N_p>$, doubly $<N_{alpha}>$ and multiply $<N_F>$ charged PFs in interactions of $^{32}$S-Em at 3.7A GeV are displayed in Table 2. The corresponding data for different projectiles ($^4$He, $^{12}$C, $^{16}$O, $^{22}$Ne, $^{28}$Si) at nearly the same energy are shown in the same table. The mean values of PFs at $^{32}$S 200A GeV [19] are also shown in Table 2 for comparison with the present work.

**Table 2.** Mean values of singly $<N_p>$, doubly $<N_{alpha}>$ and multiply $<N_F>$ charged PFs for different projectiles at ~3.7A GeV. The corresponding data of $^{32}$S at 200A GeV are displayed as well.

| Projectile | Energy (A GeV) | $<N_p>$ | $<N_{alpha}>$ | $<N_F>$ | Ref. |
|---|---|---|---|---|---|
| $^4$He | 3.7 | 0.61 ± 0.02 | 0.07 ± 0.01 | 0.00 ± 0.00 | 9 |
| $^{12}$C | 3.7 | 1.20 ± 0.04 | 0.68 ± 0.03 | 0.18 ± 0.01 | 7 |
| $^{16}$O | 3.7 | 1.45 ± 0.03 | 0.74 ± 0.02 | 0.33 ± 0.01 | 7 |
| $^{22}$Ne | 3.3 | 1.55 ± 0.02 | 0.80 ± 0.01 | 0.48 ± 0.01 | 7 |
| $^{28}$Si | 3.7 | 2.54 ± 0.05 | 1.08 ± 0.03 | 0.56 ± 0.01 | 7 |
| $^{32}$S | 3.7 | 2.00 ± 0.09 | 1.73 ± 0.08 | 0.48 ± 0.02 | This work |
| $^{32}$S | 200 | 2.28 ± 0.15 | 1.75 ± 0.08 | - | 19 |

From this table, in all interactions of different projectiles with emulsion nuclei at 3.7A GeV, the average multiplicities of all the different charged PFs ($<N_p>$, $<N_{alpha}>$ and $<N_F>$) show a decrease with increasing charge of PFs. Also, for PFs with a certain charge, the average multiplicities increase with increasing projectile size.

**Table 3.** Parameters and $\chi^2$/DoF of the Gaussian fits to the MDs for singly and doubly charged PFs.

| Z = 1 PFs | Center | Width | Peak | $\chi^2$/DoF | Z = 2 PFs | Center | Width | Peak | $\chi^2$/DoF |
|---|---|---|---|---|---|---|---|---|---|
| He | 0.10 | 2.84 | 57.26 | 1.50 | - | - | - | - | - |
| C | 0.03 | 4.88 | 42.10 | 0.03 | C | 0.00 | 1.76 | 46.15 | 4.95 |
| O | 0.74 | 4.37 | 32.26 | 0.89 | O | - 0.01 | 2.03 | 52.23 | 6.95 |
| Ne | 0.27 | 4.91 | 31.09 | 0.06 | Ne | - 0.01 | 2.86 | 50.11 | 0.46 |
| Si | 1.73 | 6.05 | 20.02 | 0.18 | Si | - 0.15 | 3.94 | 41.07 | 0.32 |
| S | - 0.39 | 4.63 | 38.90 | 0.09 | S | - 0.16 | 3.63 | 46.00 | 0.21 |

The MDs of Z = 1 and Z = 2 charged PFs for all interactions in Figs. 1(a) and 1(b), respectively have been fitted well with a Gaussian distribution function. The parameters of the Gaussian fit are the central value, the width and the peak of the distribution. For all the interactions, the values of the parameters and $\chi^2$/DoF for each Gaussian fit are calculated and demonstrated in Table 3.

## 3.2 Dependence of the PFs on projectile energy

The MDs of singly and doubly charged PFs for $^{32}$S-Em collisions at the lowest (3.7A GeV) and highest (200A GeV) available energies are shown in Fig. 2 (a) and Fig. 2 (b). The probability of producing singly and doubly charged PFs decreases as the number of these fragments per event increases. Moreover, the measurements of $<N_p>$ and $<N_{alpha}>$ for the $^{32}$S projectile given in Table 2 are independent of the beam energy.

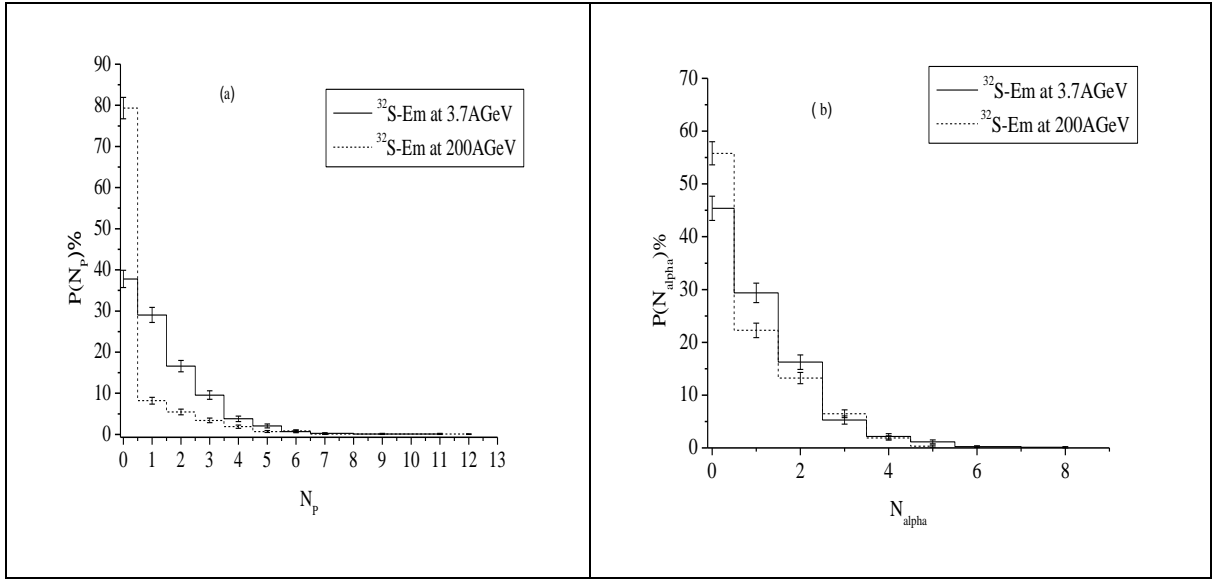

**Fig. 2.** Multiplicity distributions of a) singly and b) doubly charged PFs for $^{32}$S-Em interactions at 3.7A and 200A GeV, respectively.

The usual way of investigating the charged-particle MD and its shape, is to calculate its moments. In statistics, probability distribution functions can be characterized by the various order moments, such as the mean ($<N>$), the variance ($\sigma^2 = <(\Delta N)^2>$), the skewness (Skew $= <(\Delta N)^3>/\sigma^3$) and the kurtosis ($K = <(\Delta N)^4>/\sigma^4 - 3$), where $\Delta N = N - <N>$ [20]. Skewness and kurtosis are used to characterize the asymmetry and peakness of the MDs, respectively. For Gaussian distributions, they are equal to zero. Thus, the skewness and kurtosis are an ideal probe to demonstrate the non-Gaussian fluctuation feature [21].

In Table 4, we calculate the various order moments of the MD of the singly and doubly charged PFs produced in the $^{32}$S-Em collisions at 3.7A and 200A GeV. We also calculate the scaled variance ($\omega = \sigma^2/<N>$) to measure the multiplicity fluctuation in the production of charged PFs. The weak bound deduced from Table 4 is that the distributions of Z = 1 and Z = 2 charged PFs at both energies cannot be wider than Poissonian, i.e., $\omega \leq 1$. The results in Table 4 also show that the distributions of singly and doubly charged PFs at both energies are positively and highly skewed, meaning that the right tail of each distribution is longer than the left tail. Additionally, it seems from Table 4 that kurtosis in the case of singly charged PFs has a stronger energy dependence than that of doubly charged PFs. The positive values of the kurtosis in the case of singly charged PFs at 3.7A GeV and in the case of doubly charged PFs at both energies indicate the possibility of a slightly platykurtic distribution (kurtosis < 3) [20] while, the positive value of kurtosis in the case of singly charged PFs at 200A GeV indicates the possibility of a strongly leptokurtic distribution (kurtosis > 3).

**Table 4.** Variance, skewness, kurtosis and ratio of variance to mean of charged PFs produced in $^{32}$S-Em interactions.

| Energy (A GeV) | variance ($\sigma^2$) | skewness (Skew) | kurtosis (K) | $\sigma^2/<N>$ ($\omega$) |
|---|---|---|---|---|
| | | Singly charged PFs | | |
| 3.7 | 1.88±0.03 | 1.38±0.01 | 2.27±0.03 | 0.94±0.05 |
| 200 | 1.23±0.01 | 2.85±0.01 | 8.58±0.02 | 0.54±0.04 |
| | | Doubly charged PFs | | |
| 3.7 | 1.25±0.02 | 1.40±0.01 | 2.09±0.03 | 0.72±0.04 |
| 200 | 1.12±0.01 | 1.34±0.01 | 1.19±0.02 | 0.64±0.03 |

Figure 3 shows the variation of average number of charged PFs $<N_i>$ as a function of the projectile mass number $A_P$ in nucleus-emulsion (A-Em) interactions at ~ 3.7A GeV. This dependence is usually parameterized in the following power law form:

$$<N_i> = a_i\, A_p^{b_i} \tag{1}$$

where, $a_i$ and $b_i$ are fitting parameters and i denotes singly, doubly or multiply charged PFs.

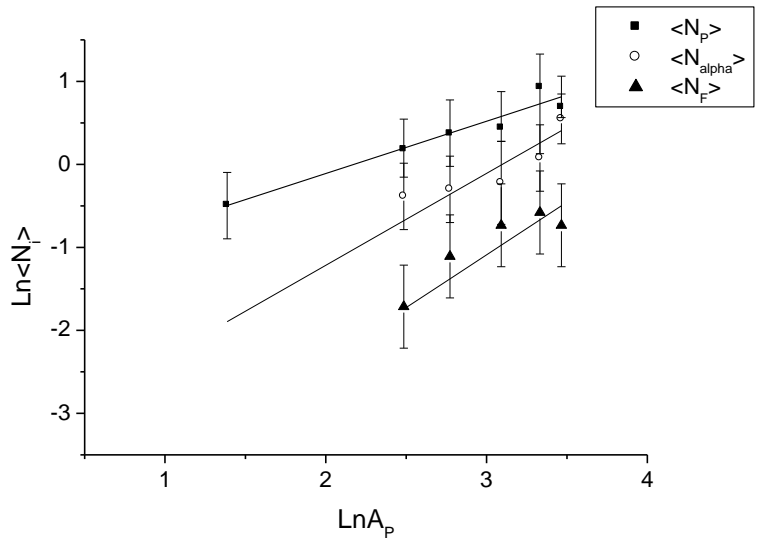

**Fig. 3.** Average multiplicities of PFs $\langle N_i \rangle$ as a function of the projectile mass number $A_P$ at 3.7A GeV.

The straight lines in Fig. 3 represent the power law fitting. The fitting parameter values for different charged PFs are given in Table 5. The values of $\chi^2$/DoF are shown in the same table. From Table 5, $\langle N_p \rangle$ is proportional to $A_P^{2/3}$ while, both $\langle N_{alpha} \rangle$ and $\langle N_F \rangle$ are nearly proportional to $A_P^{4/3}$.

**Table 5.** Fitting parameter values of Eq. (1).

| $\langle N_i \rangle$ | $b_i$ | $a_i$ | $\chi^2$ / DoF |
|---|---|---|---|
| $N_P$ | 0.63±0.08 | -1.37±0.23 | 0.04 |
| $N_{alpha}$ | 1.11±0.37 | -3.43±0.92 | 0.07 |
| $N_F$ | 1.27±0.22 | -4.90±0.64 | 0.08 |

### 3.3 Dependence of the PFs on different target groups

Here, we study the dependence of the MDs of singly charged, $N_p$ and doubly charged, $N_{alpha}$ PFs on different emulsion target groups. Figures 4 and 5 respectively show the MDs of both Z = 1 and Z = 2 PFs emitted in the interactions of $^{32}$S nuclei with hydrogen H, light CNO and heavy AgBr groups of emulsion nuclei at 3.7A GeV. From these figures, there are no significant differences in the yields of Z = 1 and Z = 2 PFs, indicating that the break-up mechanism of the $^{32}$S projectile seems to be independent of the mass of the target.

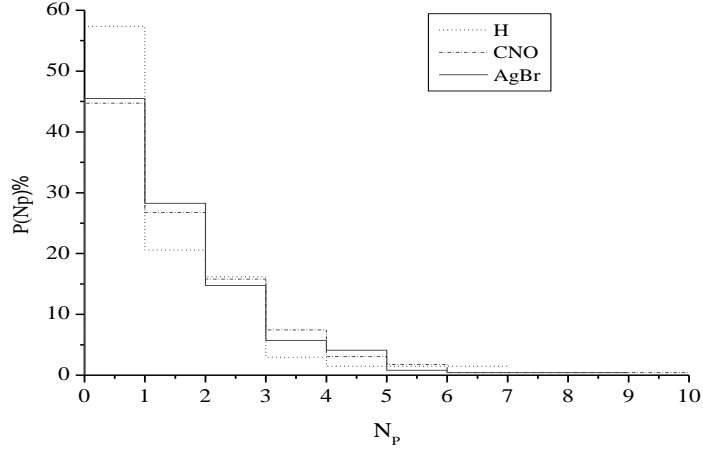

**Fig. 4.** Multiplicity distribution of singly charged PFs $N_P$ produced from the interactions of $^{32}$S 3.7A GeV with different target groups.

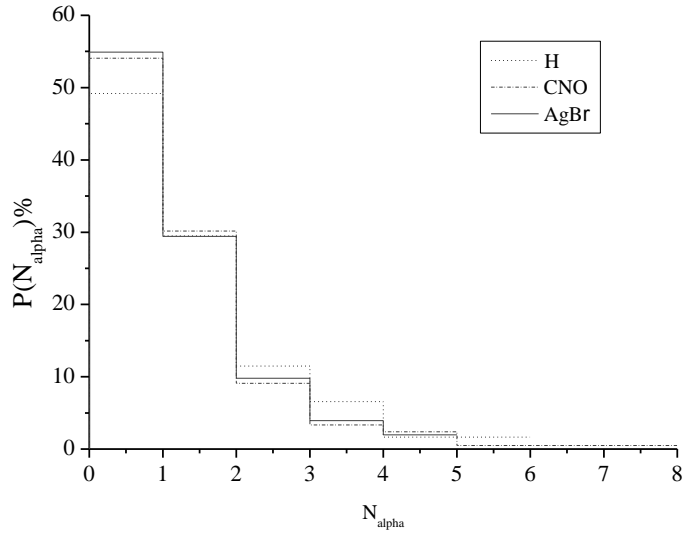

**Fig. 5.** Multiplicity distribution of doubly charged PFs produced from the interactions of $^{32}$S 3.7A GeV with different target groups.

**Table 6.** Mean multiplicity of Z = 1 PFs produced in interactions of different projectiles with the three target groups at nearly the same energy. The corresponding data of $^{32}$S at 200A GeV are shown as well.

| Projectile | Energy (A GeV) | <$N_P$> | | | Ref. |
|---|---|---|---|---|---|
| | | H | CNO | AgBr | |
| $^{12}$C | 3.7 | 1.20±0.09 | 1.38±0.07 | 1.03±0.06 | 7 |
| $^{16}$O | 3.7 | 1.38±0.06 | 1.60±0.05 | 1.35±0.04 | |
| $^{22}$Ne | 3.3 | 1.18±0.04 | 1.70±0.04 | 1.63±0.04 | |
| $^{28}$Si | 3.7 | 2.25±0.11 | 2.77±0.09 | 2.51±0.08 | |
| $^{32}$S | 3.7 | 1.76±0.19 | 2.06±0.14 | 2.01±0.13 | This work |
| $^{32}$S | 200 | 2.19±0.30 | 2.32±0.23 | 2.18±0.22 | 19 |

**Table 7.** Mean multiplicity of Z = 2 PFs produced in interactions of different projectiles with the three target groups at nearly the same energy. The corresponding data of $^{32}$S at 200A GeV are shown as well.

| Projectile | Energy (A GeV) | <$N_{alpha}$> | | | Ref. |
|---|---|---|---|---|---|
| | | H | CNO | AgBr | |
| $^{12}$C | 3.7 | 1.17±0.07 | 0.69±0.04 | 0.37±0.03 | 7 |
| $^{16}$O | 3.7 | 0.94±0.05 | 0.89±0.04 | 0.47±0.03 | |
| $^{22}$Ne | 3.3 | 0.92±0.03 | 0.92±0.03 | 0.63±0.02 | |
| $^{28}$Si | 3.7 | 1.30±0.08 | 1.16±0.05 | 0.91±0.05 | |
| $^{32}$S | 3.7 | 1.87±0.24 | 1.74±0.12 | 1.69±0.12 | This work |
| $^{32}$S | 200 | 1.85±0.16 | 1.76±0.12 | 1.66±0.12 | 19 |

Table 6 shows the mean multiplicity of Z = 1 PFs produced in interactions of $^{32}$S at 3.7A GeV in comparison with different projectiles ($^{12}$C, $^{16}$O, $^{22}$Ne, $^{28}$Si) at nearly the same energy. The corresponding values of $^{32}$S at 200A GeV are shown in the same table as well. This table shows that <$N_P$> values are almost independent of target mass $A_T$. Additionally, <$N_P$> values from $^{32}$S projectiles seem to be independent of the incident energy.

Table 7 represents the mean multiplicity of Z = 2 PFs produced in interactions of $^{32}$S at 3.7A GeV and the corresponding results from different projectiles ($^{12}$C, $^{16}$O, $^{22}$Ne, $^{28}$Si) at nearly the same energy. The corresponding values of $^{32}$S at 200A GeV are also shown in the same table. For different projectiles, the value of $<N_{alpha}>$ decreases slowly as the target mass increases while, for $^{32}$S projectiles, $<N_{alpha}>$ values within experimental errors seem to be energy independent.

## 4 Conclusion

On the basis of the experimental data containing 868 inelastic interactions of 3.7A GeV sulphur ions with emulsion nuclei, the main characteristics of 1081 singly, 821 doubly and 420 heavily charged PFs are investigated and compared with the corresponding results for other projectiles ($^{12}$C, $^{16}$O, $^{22}$Ne, $^{28}$Si) at the same incident energy and the following conclusions can be drawn:

1- The MDs of different PFs represented by the probabilities of events suggest the validity of the limiting fragmentation hypothesis.

2- The MDs of Z = 1 and Z = 2 charged PFs for all studied interactions at Dubna energy have been fitted well with a Gaussian distribution function.

3- A decrease of the scaled variance for the multiplicity fluctuations has been found, revealing that the distribution of Z = 1 and Z = 2 charged PFs at both energies (3.7A and 200A GeV) cannot be wider than Poissonian, i.e., $\omega \leq 1$.

4- The measurements of average multiplicities of the different charged PFs ($<N_p>$, $<N_{alpha}>$ and $<N_F>$) in interactions of different projectiles with emulsion nuclei at 3.7A GeV tend to decrease with increasing fragment charge. For the PFs with a certain charge, the average multiplicities increase with increasing projectile mass.

5- The value of $<N_P>$ increases with the projectile mass $A_p$ with exponent 2/3, which shows a strong correlation between the number of Z = 1 PFs emitted and the surface area of the emitting source while, $<N_{alpha}>$ and $<N_F>$ have almost the same value, both depending on $A_P$ with exponent $\approx 4/3$. In other words, the study of mean multiplicities of different charged PFs against the projectile mass shows a power-law relationship.

6- For $^{32}$S projectiles, the values of $<N_p>$ and $<N_{alpha}>$ at both incident energies (3.7A and 200A GeV) have the same values within experimental errors, indicating that the fragmentation of the projectile nucleus does not depend on the projectile energy. This behavior has been previously noticed in Ref. [19].

7. At both incident energies, events with two heavy PFs ($Z > 2$) are scarce and their fraction does not exceed 2.0 %, while the production of a single PF with $Z > 2$, which is the most probable mode of fragmentation, decreases with energy.

8. The energy dependence of lower moments (Skew, K) of $Z = 1$ and $Z = 2$ MDs is observed to indicate that the shape of MDs is positively and highly skewed, while kurtosis in case of singly charged PFs has stronger energy dependence than that of doubly charged PFs.

9. For all target emulsion groups, the values of $<N_P>$ at 200A GeV are slightly higher than that at 3.7A GeV while, the values of $<N_{alpha}>$ for all target groups are unchanged at both incident energies. So, the emission of doubly charged PFs seems to be energy independent.

10. The mean multiplicities of PFs decrease with target mass $A_T$, except for singly charged PFs where its mean multiplicity is almost independent of $A_T$.

*The authors are grateful for the guidance and advice given by Prof. Dr. A. Abdelsalam, the head of Mohamed El–Nadi High Energy Laboratory., at Cairo University, where this work has been carried out. We wish to acknowledge the staff of the high-energy lab. at JINR, Dubna, Russia, for supplying the irradiated emulsion plates.*